\documentclass[a4paper, 12pt]{article} 
\usepackage{fixltx2e,fix-cm}
\usepackage{amssymb}
\usepackage{amsmath}
\usepackage{graphicx}
\usepackage{subfigure}
\usepackage{multicol}
\usepackage{bm}
\usepackage{natbib}

\title{Basic stochastic transmission models\\ and their inference}
\date{\today}
\author{Tom Britton$^1$}

\begin{document}
\maketitle

\begin{abstract}
The current survey paper concerns stochastic mathematical models for the spread of infectious diseases. It starts with the simplest setting of a homogeneous population in which a transmittable disease spreads during a short outbreak. Assuming a large population some important features are presented: branching process approximation, basic reproduction number $R_0$, and final size of an outbreak. Some extensions towards realism are then discussed: models for endemicity, various heterogeneities, and prior immmunity. The focus is then shifted to statistical inference. What can be estimated for these models for various levels of detailed data and with what precision? The paper ends by describing how the inference results may be used for determining successful vaccination strategies. This paper will appear as a chapter of a forthcoming book entitled \emph{Handbook of Infectious Disease Epidemiology}.

\end{abstract}

\footnotetext[1]{Department of Mathematics, Stockholm University, 106 91 Stockholm, Sweden.\\ Email: tom.britton@math.su.se}




\section{Introduction}

The current chapter aims at presenting some basic stochastic models for the spread of infectious diseases in human (or animal) populations, and to also describe how to perform inference about important model parameters, such as the basic reproduction number $R_0$ and the critical vaccination coverage $v_C$. Naturally, there is some overlap, but also differences, with the current chapter and other overview papers, in particular two by the same author. However, \cite{Brit2010} has more focus on the stochastic analysis of models and only briefly touches upon inference procedures, and \cite{BritGiard2016} describe briefly many different inferential aspects with extensive references to the literature. In the current paper we focus on basic models and try to be more self-contained, more complex and realistic models are treated in later chapters of the book. There are of course numerous papers dealing with this type of inference. Two recent books on the topic are \cite{Beck2015} and \cite{DHB12}, the latter also doing extensive modelling and being more theoretical.

The mathematical/statistical models describe the spread of a transmittable disease. What makes such diseases different from other diseases, both regarding the mathematical analysis but also in reality, is that transmittability implies that the health status of different individuals will be \emph{dependent}, as opposed to other diseases where the occurence of diseases in different individuals happen independently. These dependencies make the mathematical treatment, as well as the statistical analysis, more involved, as we will see. We will present some simple models and only briefly discuss extensions towards more realistic models, and the presented inference procedures will focus on estimation of basic parameters.

The rest of this chapter is structured as follows. In Section \ref{Sec_model} we define the basic models to be used, and in the next section we discuss some model extensions. In Section \ref{Sec_inference} we present the main inference procedures, for a couple of different types of data. In Section \ref{Sec_prevention} we study effects of preventive measures put in place before or during an outbreak, and how such effects may be estimated from previous outbreak data.

\section{The standard stochastic SIR epidemic model}\label{Sec_model}

The class of models we analyse are where individuals may be classified into three classes: Susceptibles (individuals who have not experienced the disease but who are susceptible to infection), Infectives (individuals who have been infected and may transmit the disease onwards), and Recovered (who can no longer transmit the disease and who are immune to the disease). Such models are called SIR models from the three classes and how individuals may move between the three states. If individuals who get infected first enter a latent state before becoming infectious, the models are called SEIR model where "E" stands for Exposed but not yet infectious. If immunity is not permanent but wanes, the model would be called an SIRS model indicating the non-transient nature of such a model.

We consider a population of size $n$, where approximations/limit results rely on $n$ being large. When we look at short term outbreaks we consider a fixed population of size $n$, whereas later, when considering endemic diseases, we let $n$ denote the average population size in a community in which individuals die and new are born.

\subsection{Definition: the Standard stochastic SIR epidemic}

We now define what we call the \emph{Standard stochastic SIR epidemic} in a fixed and closed community. Consider a comunity of size $n$ in which an SIR epidemic spreads. Initially all individuals are susceptible except one index case who is infectious. Individuals who get infected remain infectious for a random period $I$, having mean $E(I)=\iota$, and then recover. Infectious individuals have infectious contacts at rate $\beta$, each time with a uniformly chosen individual in the community. An infectious contact with a susceptible individual implies that the latter gets infected whereas other contacts have no effect. The epidemic goes on (infectious individuals having infectious contacts until they recover) until the first time $T$ when no one is infectious. Then the epidemic stops.

We let $S(t), I(t)$ and $R(t)$ respectively denote the number of susceptible, infectious and recovered, at time $t$ measured from the start of the epidemic. Since the population is fixed and closed we have $S(t)+I(t)+R(t)=n$ for all $t$. The corresponding fractions are denoted $\bar S(t)=S(t)/n$ and similarly. Whenever the dependence on $n$ is important we equip the quantities with an $n$-index. As regards to parameters, we have the infectious contact rate $\beta$ and the duration of the infectious period $I$ being a random variable. Of fundamental importance is $R_0:= \beta E(I)=\beta \iota$, and called the basic reproduction number. This is hence the average number of infectious contacts an infectious individual has during his/her infectious period. In the beginning of the outbreak and assuming a large community, all such contacts will be with distinct and susceptible individuals with high probability, so $R_0$ is the expected number of individuals an infected person infects in the beginning. It should hence not come as a surprise that a big (or major) outbreak can only happen if $R_0>1$.

\subsection{The general stochastic epidemic}

Two specific choices of infectious periods $I$ have received special attention in the literature. The first is where $I\sim Exp(\gamma)$ (so $\iota=1/\gamma$). This model is often called the \emph{General stochastic epidemic} (or the Markovian epidemic) and its main reason for receiving attention is that the model then becomes Markovian thus having mathematically tractable properties. In the limit as $n\to\infty$ this model corresponds to the (deterministic) \emph{general epidemic model} defined by the differential equations:
\begin{align}
s'(t) &=-\beta s(t)i(t) \nonumber \\
i'(t) &= \beta s(t)i(t) - \gamma i(t)\label{SIR-diff}\\
r'(t) &= \gamma i(t). \nonumber
\end{align}
For this model $R_0=\beta/\gamma$ and it is seen that, starting with $s(0)=1-\epsilon$, $i(0)=\epsilon$ and $r(0)=0$ for some small $\epsilon>0$, $i(t)$ is initially increasing if and only if $R_0>1$. One difference between this deterministic general epidemic and the stochastic general epidemic is that the deterministic model will surely have an outbreak infecting a substantial community fraction when $R_0>1$, whereas in the stochastic setting starting with a small \emph{number} of infectives, a major epidemic \emph{can} happen, but the epidemic may as an alternative still die out infecting only few individuals. So, in the stochastic setting there could be a minor outbreak with a certain probability and a major outbreak with the remaining probability. 

\begin{figure*}[ht]
\begin{center}
\includegraphics[scale=0.65]{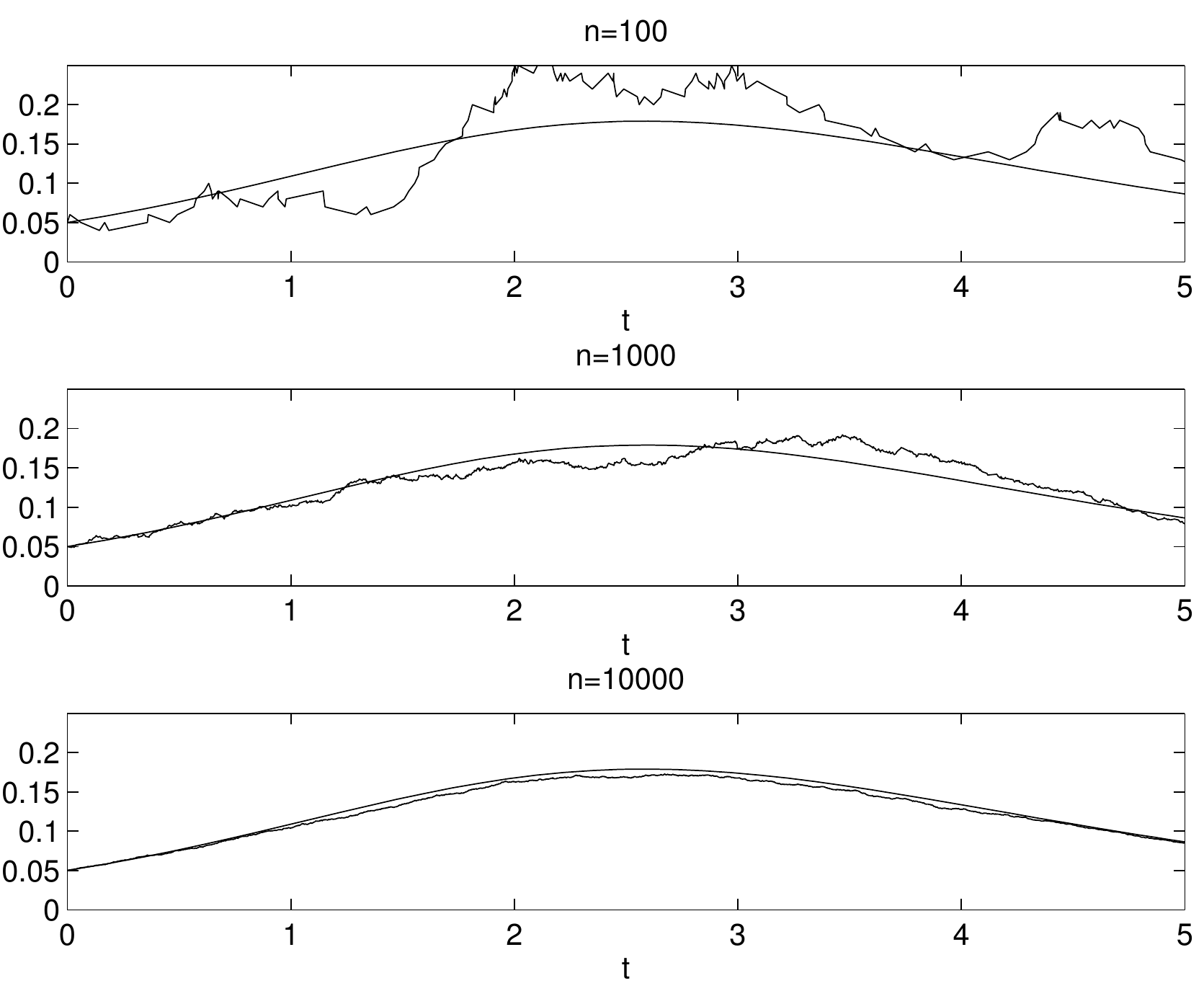}
\end{center}
\caption{ Plot of $\bar I_n(t)$ (for $n=100$, 1000 and 10~000) and its deterministic limit $i(t)$ against $t$. Parameters are $\beta=2$, $\gamma=1$ (e.g.\ weeks as time unit and average infectious periods of one week), so $R_0=2$. }\label{fig-barI}
\end{figure*}

In Figure \ref{fig-barI} we have plotted $\bar I_n(t)$ for a few different $n$, and its deterministic counterpart $i(t)$, starting with 5\% infectives thus assuring a major outbreak also in the stochastic setting. It is seen that the stochastic curve agrees better with the deterministic counterpart the larger $n$ is.

\subsection{The Reed-Frost epidemic and chain-binomial models}

The second choice of  infectious period which has received specific attention is where $I\equiv \iota$, i.e.\ where the infectious period is non-random and the same for all individuals, a model called the continuous time Reed-Frost epidemic. This model has received special attention also for mathematical rather than epidemiological reasons. One probabilistic advantage with this model is that when the infectious period is non-random, then the events for an infectious individual to infect different other individuals become \emph{independent}. When the infectious period is random this does not hold: if the infective infects another individual this indicates that most likely the infectious period was long, and this increases the risk to infect another individual. But in the Reed-Frost epidemic these events are independent, so an infective has independent infectious contacts with each other individual, and these contact probabilities all equal $p=1-e^{-\beta\iota/n}\approx \beta\iota /n$ (the contact rate to a specific other individual equals $\beta/n$). 

If individuals are latent for a period prior to the constant infectious period, and assuming the the latent period is long and the infectious period is short, then the new infected people will appear in ''generations'', something which can actually even be observed during early stages of outbreaks. This is then called the discrete-time version of the Reed-Frost epidemic. Anyway, then a susceptible individual escapes infection in generation $k+1$ if he/she avoids getting infected from each of the infected people of the previous generation, so this happens with probability $(1-p)^{i_k}$, where $i_k$ denotes the number of individuals who got infected in generation $k$. The probability to get infected is the complimentary probability $1-(1-p)^{i_k}$. This is true for all individuals who were susceptible after generation $k$ and the infection events are independent between different pairs of individuals (due to constant infectious period). As a consequence, if there are $i_k$ individuals getting infected in generation $k$ and $s_k$ remaining susceptible, then it follows that 
$$
I_{k+1}\sim Bin(s_k,\ 1-(1-p)^{i_k} )\ \text{and}\ S_{k+1}=s_k-I_{k+1},
$$ 
where $Bin(n,p)$ denotes the binomial distribution with parameters $n$ and $p$.

We can use this iteratively over different generations to compute the probability of an entire outbreak in terms of generations. As a samll example, suppose that we want to compute the probability that in a community of 10 individuals and starting with one infectious and nine susceptibles, we want to compute the probability that first 2 got infected, then 3 followed by 1, and then noone more. This means that we have $(i_0=1, s_0=9)$ followed by $(i_1=2, s_1=7)$, $(i_2=3, s_2=4)$, $(i_3=1, s_1=3)$ and $(i_4=0, s_4=3)$. The probability for this outbreak chain is given by
$$
\binom{9}{2}p^2(1-p)^7  \binom{7}{3}(1-(1-p)^2)^3((1-p)^2)^4  \binom{4}{1}(1-(1-p)^3)^1((1-p)^3)^3  \binom{3}{0}p^0(1-p)^3.
$$
This should explain why the discrete time version of the Reed-Frost model is often referred to as a \emph{chain binomial model}. It is possible to think of other chain binomial models (e.g.\ where the infection probabilities are different or there are different types of individuals) but the discrete time Reed-Frost model is by far the most well studied chain binomial model. The final size probabilities can in principle be determined by summing the different chains given a specified final size, but for more than, say 5, infected people there are to many chains giving such a final size thus making this approach of less practical use.

It is worth pointing out that the time-continuous Reed-Frost model that we started with in fact gives the same final outcome probabilities as the discrete time Reed-Frost (having the same $p$). The order in which individuals get infected, and by whom, differ in the two models, but the same number of individuals will ultimately get infected. For this reason the two models are sometimes used interchangeably.

\subsection{Asymptotic results}\label{Sec-asymptotics}

We now present some results for the standard stochastic SIR epidemic valid for large $n$. All the results can be proven to hold as limit results when $n\to\infty$.

As mentioned earlier, in the beginning of an outbreak in a large community, an infectious individual will have all its infectious contacts with distinct individuals who are susceptible. An infective will hence infect new individuals at constant rate $\beta$ during the infectious period $I$, and people he/she infects will do the same and independently. This then satisifies the definition of a continuous-time branching process, where individuals give birth (i.e.\ infect) at rate $\beta$ during their life-span (infectious period) $I$. 

The mean of the offspring distribution is given by $R_0=\beta E(I)=\beta \iota$. It is known that if $R_0\le 1$, then the branching process (i.e.\ epidemic) can never take off, and just a small number of individuals will ever get born (be infected). If however $R_0>1$, then the epidemic may take off infecting large number of individuals. In the beginning of the outbreak, each individual infects a random number $X$ new individuals, and given the duration of the infectious period $I=s$, then the number of infections is Poisson distributed with mean parameter $\beta s$ (the infection rate multiplied by the duration). Without conditioning on the infectious period, the number of infections is henced what is called a mixed Poission distribution $X\sim MixPoi(\beta I)$ where $I$ is random following the distribution specified by the model. For the continuous time Reed-Frost model $X\sim Poi(\beta \iota)$ since $I\equiv \iota$ is non-random, and for the Markovian SIR where $I\sim Exp(\gamma)$ (having mean $\iota=1/\gamma$) it is not hard to show that $X\sim Geo (\gamma/(\beta+\gamma))$.

From branching process theory we conclude the following:

\vskip.1cm
\noindent a) An epidemic can take off if and only if $R_0=\beta E(I)>1$. 

\vskip.1cm\noindent b) If $R_0>1$, the probability $\pi$ that the epidemic takes off equals the unique strictly positive solution to the equation $1-\pi=\rho(1-\pi)$, where $\rho (s)=E(s^X)$ and $X\sim MixPoi (\beta I)$ meaning that $X$ given $I=s$ is $Poi (\beta s)$ and $I$ follows the specified ditribution defined in the model. For the Reed-Frost model this equation becomes $1-\pi=e^{-R_0 \pi}$ and for the Markovian SIR the solution is explicit and equals $\pi=1-1/R_0$.

\begin{figure*}[ht]
\begin{center}
\includegraphics[scale=0.3]{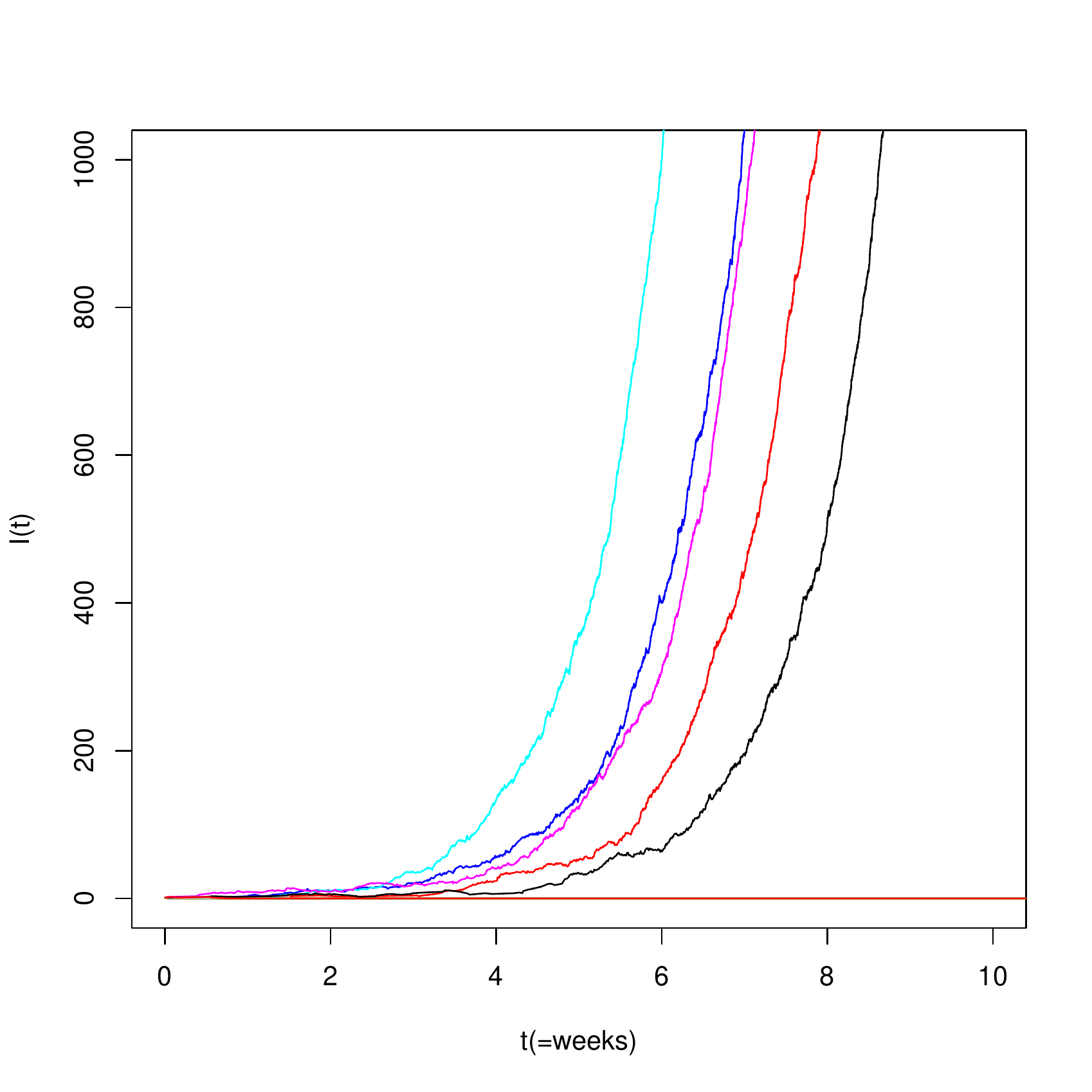}
\includegraphics[scale=0.3]{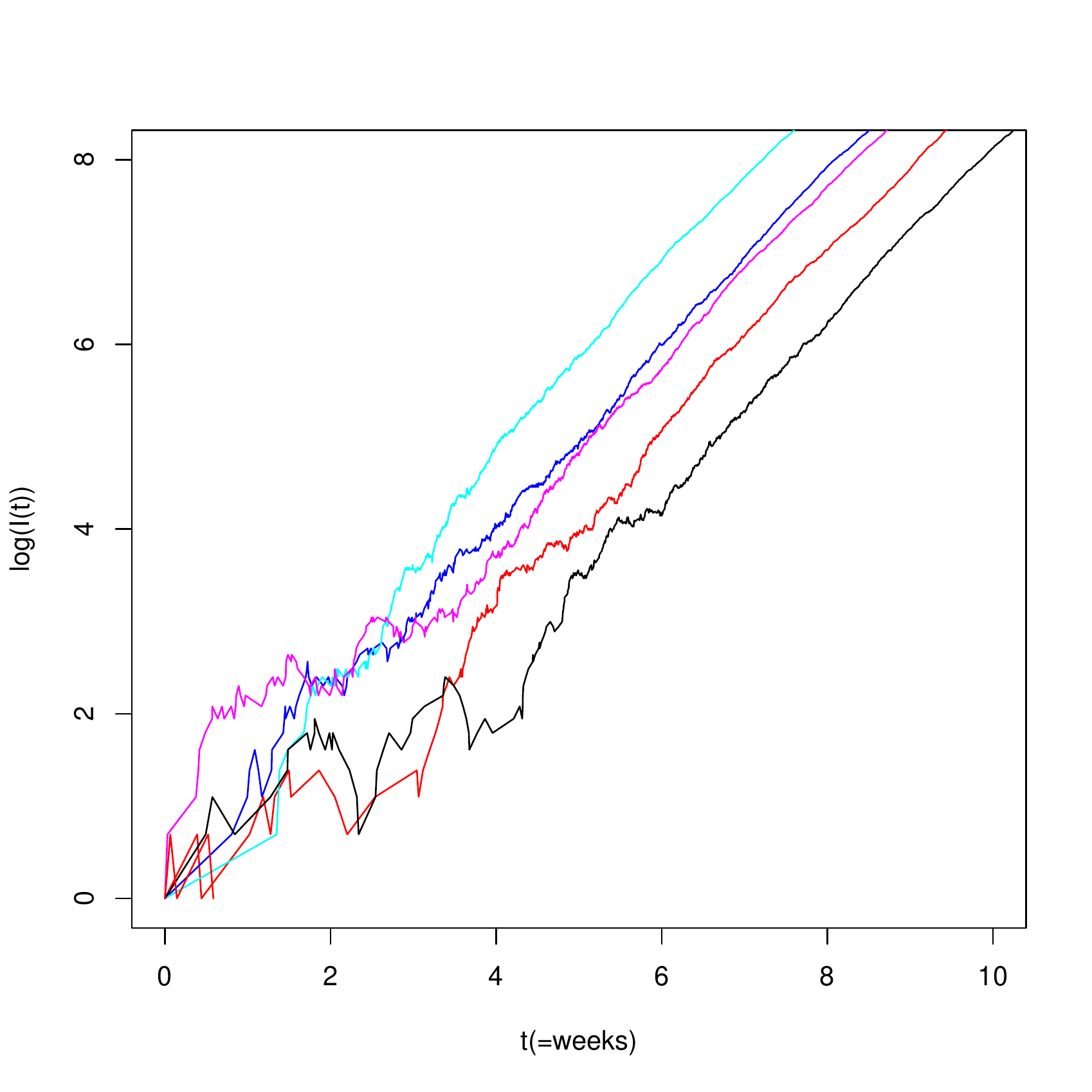}
\end{center}
\caption{Plot of $I(t)$ during the initial epidemic stage for 10 simulations, original as well as log-scale (for the ones that take off). The population size is $n=100~000$ so depletion of susceptibles have hardly started when at most 1000 individuals have been infected. Five of the simulations die out quickly whereas the remaining take off, having different initial delays before taking off. The original scale shows the exponential growth which is made even more evident on the log-scale where the growth is linear. The model parameters are $\beta=2$ and $\gamma =1$ (hence one week infectious period and $R_0=2$). The model predicts an exponential growth rate of $\rho=\beta-\gamma=1$ which corresponds to a linear growth with coefficient 1 on the log-scale (agreeing with the slopes of the lines). }\label{fig-exp-growth}
\end{figure*}

\vskip.1cm\noindent c) If the epidemic takes off (hence assuming $R_0>1$), then the number of infectives $I(t)$ at time $t$ grows exponentially in $t$: $I(t)\sim e^{\rho t}$, where $\rho $ is the so-called Malthusian parameter being the unique solution to the equation $\int_0^\infty e^{-\rho t}\beta P(I>t)dt=1$ (see Figure \ref{fig-exp-growth} for an illustration). 

\begin{figure*}[ht]
\begin{center}
\includegraphics[scale=0.5]{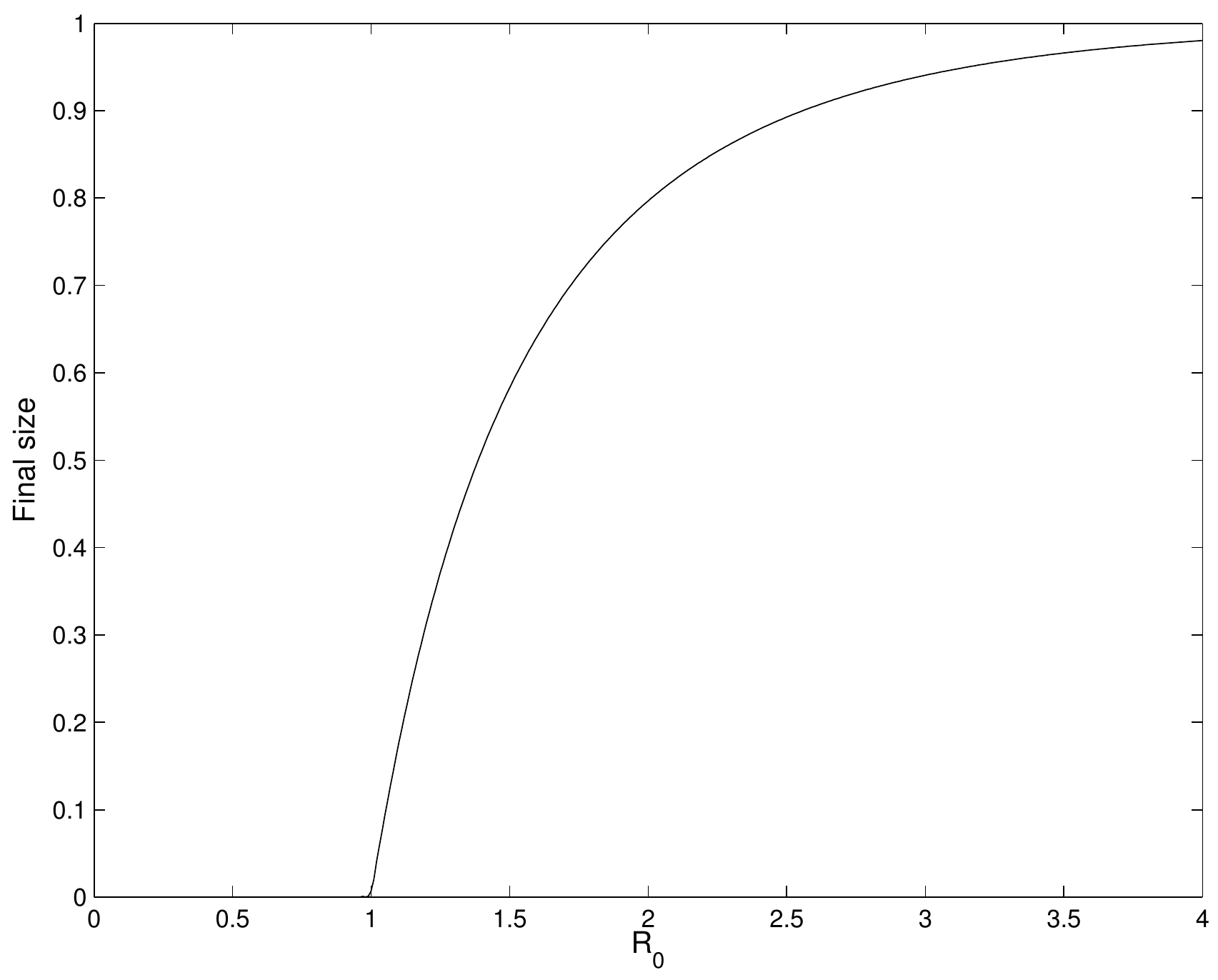}
\end{center}
\caption{The final fraction getting infected in case of a major outbreak as a function of $R_0$ (for $n\to\infty$). }\label{fig-fin-size}
\end{figure*}

\vskip.3cm If the epidemic takes off, the fraction of individuals being susceptible will start decaying so someone who gets infected will then infect fewer individuals because some of the infectious contacts will be "wasted" on already infected people. This explains why the branching process approximation, which assumes all individuals infect according to the same rules, then breaks down. It is still possible to derive approximately how many individuals that will get infected. One way to do this is by analysing the differential equations defined in Equation (\ref{SIR-diff}). By manipulating these equations it can be shown that when $t\to\infty$ and the initial fraction infectives is small and the rest are susceptible, then $r(\infty)=1-s(\infty)$, and $s(\infty)$, the fraction avoiding infection during the outbreak, is given by the positive solution to $s(\infty)=e^{R_0(1-s(\infty))}$. This equation may equivalently be expressed in terms of $r(\infty)=1-s(\infty)$: 
\begin{equation}
1-r(\infty)=e^{R_0r(\infty)},\label{fin-size}
\end{equation}
the so-called final size equation. In Figure \ref{fig-fin-size} we plot the final size $r(\infty)$ as a function of $R_0$, a solution which has to be obtained numerically.

This result is true irrespective of the distribution of the infectious period $I$ as long as $\beta E(I)=R_0$. From b) above we see that the outbreak probability for the Reed-Frost model is the same as the final size equation, so for this particular model the probability of a major outbreak (starting with one infective!) equals the final fraction getting infected in case of a major outbreak. As two numerical examples, if $R_0=1.5$ we have $r(\infty )=0.583$ so approximately 60\% will get infected if an outbreak takes place in a community without any immunity, and $r(\infty )=0.98$ if $R_0=3$.

For any finite $n$ in the stochastic setting, the ultimate fraction getting infected will of course not be exactly identical to $r(\infty)$, there will be some random fluctuations. These will however be of order $1/\sqrt{n}$, so close to negligible in large populations (in fact the randomness has been proven to be Gaussian with an explicit standard deviation which we make use of later).

In the next section we will discuss some extentions of this standard stochastic epidemic model. Here we end by emphasizing that the most important parameter $R_0=\beta E(I)$ depends both on the disease agent but also on the community under study. This can be made more explicit by writing $\beta = c\cdot p$, so $R_0=c\cdot p\cdot E(I)$, where $c$ is the rate at which individuals have close contact with other individuals, $p$ is the transmission probability for such a contact given that one individual is infectious and the other is susceptible, and $E(I)$ is the mean infectious period. Then $p$ and $E(I)$ depend on the disease agent whereas $c$ depends on the community and how frequently people have contact.

\section{Model extensions} \label{Sec_extensions}

\subsection{Including demography giving rise to endemicity}\label{Sec_endemic}

In the model defined in the previous section it was assumed that the community was fixed and closed. Such an approximation works well if considering a short term outbreak (e.g.\ influenza outbreak) taking place over a few months.

If our interest instead concerns diseases staying in the community for longer periods, like with many childhood diseases, then such an approximation is not adequate. Then we should allow for new individuals entering the community and old people leaving the community (e.g.\ by dying). Such a stochastic model can be achieved by adding a random, but with constant average rate, influx of new suscerptible individuals, and assuming that each individual dies at rate $\mu$ to the Markovian SIR model defined earlier. If we want the population size to fluctuate around $n$ this is achieved by setting the rate at which new susceptible individuals enter the community equal to $\mu n$. So, by adding influx at rate $\mu n$ and that people die at rate $\mu$ (independent of disease state) to the standard stochastic epidemic we get a simplest possible model suitable for studying endemic diseases giving life-long immunity. The corresponding defining set of differential equations for a deterministic model is given by
\begin{align}
s'(t) &=\mu -\beta s(t)i(t)-\mu s(t) \nonumber \\
i'(t) &= \beta s(t)i(t) - \gamma i(t)-\mu i(t)\label{SIR-demo-diff}\\
r'(t) &= \gamma i(t)-\mu r(t). \nonumber
\end{align}
For this model infectives have infectious contacts at rate $\beta$ until they recover or die, so now $R_0=\beta/(\gamma+\mu)$. As before, the disease will go extinct quickly if $R_0\le 1$ whereas an endemic level can be obtained if $R_0>1$. This endemic level can be obtained by setting all derivatives above equal to 0 and solving the equations. The result is 
\begin{equation}
(\tilde{s},\tilde{i},\tilde{r})= \left( \frac{1}{R_0},\ \epsilon \frac{R_0-1}{R_0}, \ 1- \frac{1}{R_0} - \epsilon \frac{R_0-1}{R_0} \right), \label{end-level}
\end{equation}
where $\epsilon=\gamma^{-1}/(\mu^{-1}+\gamma^{-1})$ is the ratio of the (average) infectious period and life-length; usually a very small number.

It is worth pointing out that the stochastic model, as well as the limiting deterministic model defined by Equation (\ref{SIR-demo-diff}), assume that the infectious period and also life-length distributions are exponentially distributed. There are extensions to more realistic scenarios but we omit them here.

\subsection{Heterogeneities}\label{Sec-heterogeneities}

The stochastic epidemic models defined above, as well as the deterministic counterparts, have all assumed a community consisting of identical individuals that mix uniformly at random with each other. Reality is of course more complicated. There are usually different types of individuals being different in terms of how susceptible they are, how much contact they have with others, and how infectious they become in case of infection. In what follows we refer to such differences as \emph{individual heterogeneities}. There is also another type of heterogeneity which concerns \emph{whom} individuals have contact with. This latter feature concerns the \emph{social structure} in the community and the fact that usually individual meet more regularly with certain individuals and much less with the remaining majority.

The individual heterogeneities are often dealt with by dividing the population into different \emph{types} of individual and assuming homogeneity within each type, meaning that individuals of the same type have the same susceptibility, total contact rate and infectivity. A corresponding epidemic is called a multitype epidemic model. Such a \emph{multitype epidemic model} is similar to the original model defined above, with the difference that now the rate of infecting someone depends on the type of the infector and the type of the susceptible type. As a consequence, $R_0$ is now more complicated -- the average number of individuals (of different types) an infected individual (of a specified type) infects is now a matrix of numbers. The basic reproduction number $R_0$ is then the \emph{largest eigenvalue} to this next generation matrix (e.g.\ \cite{DHB12}, Chapter 7). 

When it comes to the social structure of a community it depends on what type of disease is considered. For example, when considering influenza or related diseases it is common to consider \emph{household epidemic models} because spreading is usually higher within households than between other individuals. Sometimes also schools or day-care centers are included in the model. If interest is instead on sexually transmitted infections (STIs), then the relevant social structure is the sexual network in the community. Then so-called \emph{network epidemic models} \cite{Newm03} are often used, where the network obeys certain known characteristics of the empirical network but otherwise treated as random, and where an epidemic model is defined on the network. 

A different type of heterogeneity is where the contact rates vary with calendar time, often referred to as \emph{seasonality}. The simplest way to include such heterogeneity into a model is to let the infectious contact rate $\beta$ now depend on calendar time $\beta (t)$. Usually som type of periodic function is assumed, having one year as the natural period. Two such choices are $\beta (t)=a + b\sin(\omega + 2\pi t)$ where $a$ is the mean contact rate, $b$ is the amplitude of the seasonality and $\omega$ is the phase shift defining the time location of the yearly peak. A second choice is $\beta (t)=a$ for $t\in k+[t_1, t_2]$ for some integer $k$ and $\beta(t)=b$ otherwise. This means that $\beta(t)$ is a two step function, often reflecting school terms vs.\ summer break, the latter having lower overall contact rate.

Finally we mention heterogeneity in terms of the infectivity varying with time since infection. In the presented model it was assumed that individuals immediately become infectious upon infection and infect others at rate $\beta$ until the end of the infectious period when infectivity suddenly drops to 0. A more realistic model is to assume that the infectivity depends on the time $s$ since infection $\beta(s)$. For instance, there might be very low infectivity shortly after infection, then the infectivity picks up after a few days and remains high for some time until it starts decaying down to 0. It could also be that $\beta(s)$ is random in the sense that different individuals have different infectivity curves (this is actually the case also for the original model since the end of the infectious period is random). One special case of this more general 
model is where each individual is a first latent for a random period having no infectivity, followed by an random infectious period $I$ when the individual has infectious contacts at constant rate $\beta$, and then the individual recovers, the difference from the original model hence being a latent period prior to infectivity. Such models are called SEIR epidemic models, where "E" stands for exposed but not yet infectious. In terms of the epidemic, SEIR epidemics will result in the same final size (assuming the same $R_0$ of course) but the timing and duration of the outbreak will differ. From an inference point of view this means that extending the model in this direction is not important for final size data, but e.g.\ when data comes from the beginning of an outbreak time varying infectivity is often important to take into consideration.

\subsection{Prior immunity}\label{Sec-immunity}

In the model defined in Section \ref{Sec_model} it was assumed that initially everyone was susceptible to the disease except for one or a few index cases. 
 In empirical settings there is often some natural immunity in the community due to prior history to the disease (see Section \ref{Sec_prevention} for immunity due to preventive vaccination).

Suppose as a simple illustration that a fraction $s$ in the community are fully susceptible and the remaining fraction $1-s$ are completely immune. If the disease is then introduced by a few index cases the reproduction number is reduced from $R_0$ to $R_{E}=R_0s$ since, early on in the outbreak, only a fraction $s$ of all contacts  will result in infection. An outbreak is then possible only if the effective reproduction number $R_{E}>1$. We hence see that an outbreak is only possible if $s>1/R_0$. How many that get infected in case of an outbreak (as well as the probability for a major outbreak) can be derived analougously to the case without natural immunity. The result is that the fraction of the initially susceptible that ultimately get infected, $r_s(\infty)$, is the solution to the new final size equation
\begin{equation}
1-r_s(\infty)=e^{-R_0sr_s(\infty)}.\label{eff-fin-size}
\end{equation}
The overall fraction that get infected is hence $sr_s(\infty)$. As a numerical illustration, suppose $R_0=3$ and $s=50\%$, so only half of the community are susceptible. Then $r_s(\infty)=0.583$ so the overall fraction getting infected will be about 29-30\%. Compare this with the situation where there is no prior immunity (so $s=100\%$) when we saw earlier that 98\% get infected! These differences are also very important when making inference as we shall see later: neglecting prior immunity when estimating $R_0$ can lead to dramatic \emph{underestimation} of $R_0$ if not taken into account!

\section{Statistical inference} \label{Sec_inference}
 
In the previous sections we have introduced some basic epidemic models and discussed some extensions towards more realistic models. What follows now, which is the main focus of the the whole book, concerns how to make inference about model parameters after having observed an outbreak taking place.

Stochastic epidemic modelling is concerned with deriving likely outcomes given some parameter set-up. Epidemic inference goes in the opposite direction: which parameters are best in agreement with an observed outcome? This should explain why knowing some results from stochastic epidemic modelling helps when making inference.

How to make inference depends on two things: what model is considered, and what type of data that is available for making inference. In the current section our emphasis is the standard stochastic epidemic model, but we discuss two different types of data: the final size, when we observe how many that were infected at the end of the outbreak, and the situation where we also have some temporal information. We start with the former.

\subsection{Inference based on final size}

Consider a community of size $n$ and suppose that prior to the outbreak the fraction $s$ were susceptible to the disease and the rest were immune to the disease. After the outbreak has taken place we observe that a fraction $\tilde r_s$ of the initally susceptibles were infected during the outbreak. This means that we know the population size $n$ and the initial fraction immune $1-s$, and our data observation is the fraction $\tilde r_s$ among the susceptibles who got infected. 

If we only observe the final size we cannot estimate any rates or durations, so $\beta$ and $E(I)$ cannot be estimated separately, only their product $R_0=\beta E(I)$.

From Equation (\ref{eff-fin-size}) we know that $\tilde r_s$ should approximately equal the solution of this equation. A very natural estimator is hence to rewrite (\ref{eff-fin-size}) having $R_0$ on one side and to estimate $R_0$ by inserting the observed fraction $\tilde r_s$ infected. This gives the following estimator:
\begin{equation}
\hat R_0 = \frac{-\ln(1-\tilde r_s) }{s\tilde r_s}.\label{hat-R_0}
\end{equation}
As mentioned earlier it has been shown that the final fraction infected is Gaussian having mean as defined by (\ref{eff-fin-size}) and with explicit standard deviation of order $1/\sqrt{n}$. This result together with the so-called $\delta$-method (e.g.\ \cite{Rice2006}, Ch 4) can be used to obtain a standard error for the estimate $\hat R_0$. The result is
\begin{equation}
s.e.(\hat R_0 )= \frac{1}{\sqrt{ns}}\sqrt{\frac{1+c_v^2(1-\tilde r_s )\hat R_0^2s^2}{ s^2 \tilde r_s (1-\tilde r_s )}},
\end{equation}
where $c_v:=\sqrt{V(I)}/E(I)$ denotes the coefficient of variation of the infectious period. For the Reed-Frost epidemic $c_v=0$ and for the Markovian SIR $c_v=1$ and most often when estimated $c_v$ lies somewhere inbetween these two values. If unknown, a conservative estimate is hence to set $c_v=1$. Recall that $s$ is the initial fraction susceptible which is assumed to be known. If there is no natural immunity $s=1$. 

The inference presented above assumes that all infected cases are observed, meaning that there is no under-reporting. In reality there is of course under-reporting in that only some fraction $\pi$ of all cases are reported. However, if all we observe is the  fraction of reported cases among the initially susceptible, $r_s^{(rep)}$, it is impossible to deduce how many unreported cases there were. As a consequence, what fraction $\pi$ of all cases that are reported has to be inferred in some other way. Having done this we immediately have an estimated of the true fraction infected among the initially susceptible: $\hat r_s=r_s^{(rep)}/\hat \pi$. This estimate can then be used in the above expression to obtain an estimate of $R_0$. The uncertainty of the estimate increases some, how much depends on the uncertainty of the estimate $\hat \pi$ -- a standard error can be obtained using the $\delta$-method.

\subsection{Inference based on temporal data}

Quite often there is temporal information available from an outbreak, weekly reported number of cases being the most common. The date at which an infected individual is reported is typically when he or she starts showing symptoms, or rather a few days after this when a test is taken at a clinic (and later confirmed as positive). It is not always clear how this time relates to the time of infection and time of recovery, and this will depend on the disease in question. A common way to proceed is to assume that the reporting date approximately equals the recovery date (perhaps the individual receives some treatment reducing infectivity and also the ilness usually have the effect of reducing social activity). With such an assumption, and neglecting that the recovery time is often truncated to week, we hence observe $R(t)$ during some time interval $[t_0, t_1]$, often the start and end of the outbreak. There exists inference procedures for this type of data, here we simplify the situation by assuming that we also observe the infection times of individuals, thus saying that we observe $(S(t), I(t),R(t))$ for $t\in [t_1, t_2]$ together with observing the infectious periods $I_1,\dots ,I_k$ for all individuals who also recover during the period. This is the data used for inference in this section. The more likely data, observing times of diagnosis rounded to nearest week, is hence less informative but on the other hand more informative as compared to final size data considered in the previous section.

The parameters we want to make inference about are: $R_0=\beta E(I)$, and possible also the infectious contact rate $\beta$ and properties of the infectious period separately. In fact, the main advantage from having temporal information lies in the possibility to infer not only $R_0$ but the the other parameters separately, and also to be able to check model fit better. 

To estimate $R_0$ from this temporal data can be done by only using the final size data and using methods of the previous section. This estimate can be improved slightly by inserting the separate estimates obtained below: $\hat R_0=\hat \beta \hat E(I)$. For standard errors we refer to \cite{DHB12}, Sec. 5.4.2.

To infer parameters of the infectious period is straightforward, since we have i.i.d.\ observations $I_1,\dots I_k$ of the infectious period. So, for example we can estimate the mean nonparametrically by $\hat E(I)=\bar I$, the mean length of the infectious periods.

With regards to the transmission parameter $\beta$, it should be clear from Equation (\ref{SIR-diff}) that a sensible estimator for $\beta$ is obtained by integrating both sides of the top equation of (\ref{SIR-diff}), and replacing the deterministic fraction with the corresponding observed fractions: 
\begin{equation}
\hat \beta = \frac{\bar S(0)-\bar S(t)}{\int_0^t\bar S(u)\bar I(u)du}. 
\end{equation} 
In fact, $\bar S(0)-\bar S(t)- \int_0^t\beta \bar S(u)\bar I(u)du$ is a so-called \emph{martingale} which can be used to show that the estimator $\hat \beta$ is consistent and asymptotically normally distributed with an explicit standard error. For details we again refer to \cite{DHB12}, Sec. 5.4.2.

As mentioned above, another advantage with having temporal data is to check model fit. For example, one could plot the deterministic curves of Equation \ref{SIR-diff} with $\hat\beta$ and $1/\bar I$ replacing $\beta$ and $\gamma$ and compare these curves with the corresponding observed curves $(\bar S(t), \bar I(t), \bar R(t))$. If there is big discrepancy it could be that some heterogeneity has high influence on the observed epidemic which hance should be investigated further. 

Like always, the problem of underreporting is an issue also here. If it is anticipated that underreporting is substantial, then this should be estimated somehow, preferably using other sources of information (there is ongoing research aiming at estimating the underreporting fraction $\pi$ using only reported data, e.g.\ \cite{Lev2014}, the conclusion seems to be that it is problematic.

\subsection{Inference from emerging outbreaks}\label{Sec-emerging}

In the previous section the focus was on observing a complete outbreak also having some temporal information. As mentioned earlier, a complicating factor with inference for infectious diseases are the strong dependencies between infection events clearly manifested in that the rate of having infectious contact is $\beta$, but the rate of infecting new people is $\beta \bar S(t)$, since only contacts with susceptibles (which happens with probability $\bar S(t)$ at time $t$) result in infection. 

\begin{figure*}[ht]
\begin{center}
      \includegraphics[width=0.45\textwidth, height=0.6\textheight, angle=90]{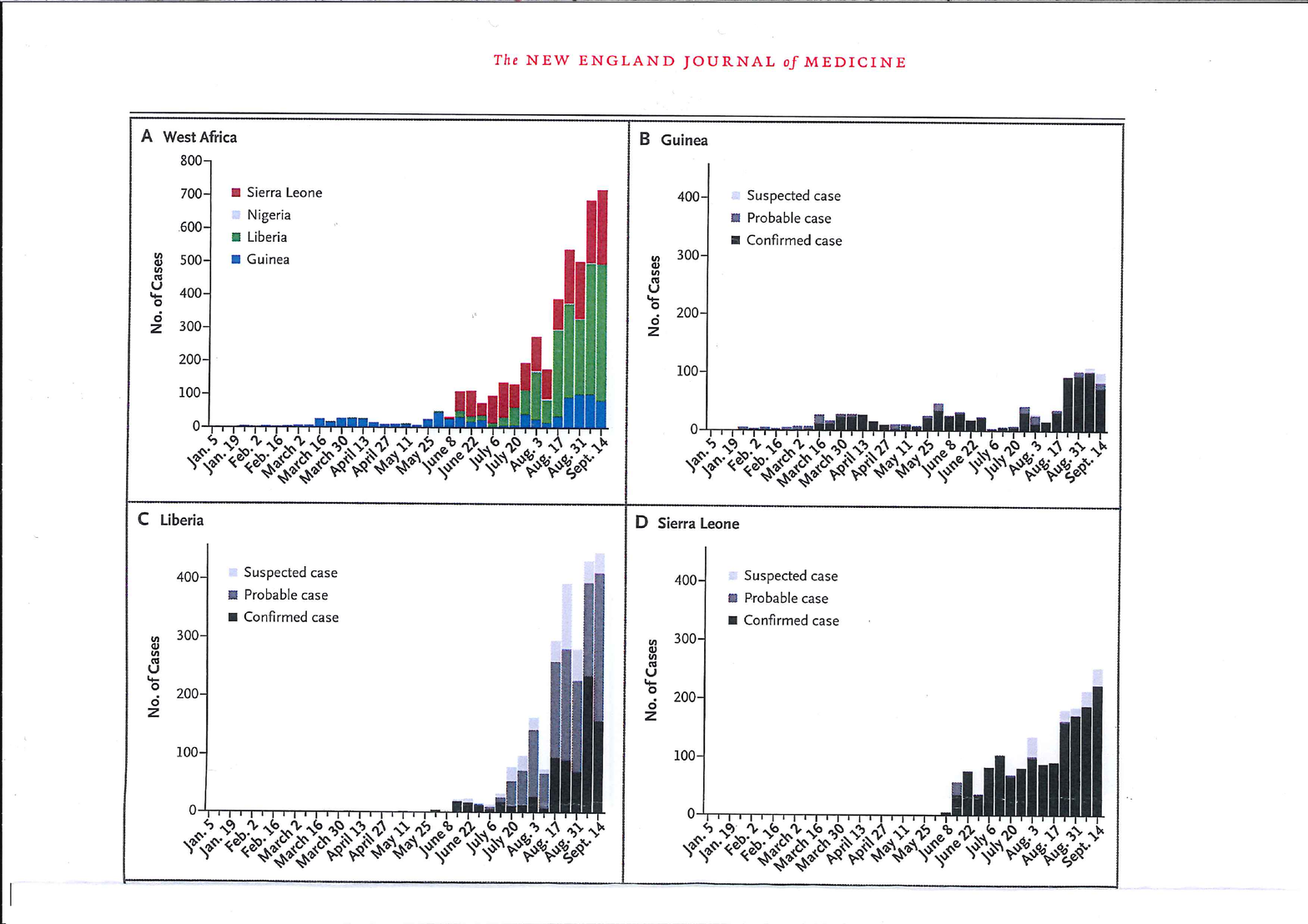} 
\end{center}
\caption{Reported number of cases of Ebola during the 2014-15 outbreak.}\label{fig-emerging}
\end{figure*}

During the early stage of an outbreak, say before 1\% have been infected, this dependence is close to negligible; so with good approximation we can assume that individuals infect new people independently (remember that we consider a homogeneously mixing community; when spreading is high within households this does not hold true). When individuals infect new people independently the epidemic model behaves like a \emph{branching process}, which we will make use of later.
In the current section we consider this type of simpler (but still hard!) situation, a suitable approximation when observing an emerging epidemic outbreak (during which typically $R(t)$ grows exponentially with rate $\rho$ say, cf. Section \ref{Sec-asymptotics}). In Figure \ref{fig-emerging} the reported number of Ebola cases during the beginning of the 2014-15 outbreak are plotted, for each of the three countries separately and together (the latter showing a clear exponentially growing behavior).

Suppose hence that we observe the number of reported cases $R(t)$, also called the reported incidence, from the start $t=0$ up until some time $t=t_1$. Using previous notation we hence observe $R(t),\ 0\le t\le t_1$ during the beginning of an outbreak meaning  that the overall fraction infected $\bar R(t_1)$ is still small (in Figure \ref{fig-emerging} much less than 1\% have been infected). Questions of interest are: what is $R_0$, how fast does the epidemic grow, and how many will eventually get infected (with or withour some specified preventive measures put in place)? 

We start with the easiest question which concerns the exponential growth rate $\rho$. Since growth is exponential and the depletion of susceptibles is still negligible, taking logarithms of the incidence and performing regression gives a simple and good estimate of $\rho$.

The remaining questions, what is $R_0$ and how many will eventually get infected, let's say without preventive measures, is harder. From observing only the initial growth (e.g.\ Figure \ref{fig-emerging}) it is in fact impossible to say anything more than that $R_0>1$ and that a substantial fraction will get infected. This should be clear from the following example. Consider two different diseases, both having $R_0=1.5$ (and assuming no prior immunity) but one having average infectious period 3 days and the other having one week average infectious period and lower daily infectivity. Since $R_0=1.5$ we know from Section \ref{Sec-asymptotics} that close to 60\% will get infected for both diseases. However, from the fact that the first disease has shorter infectious period and hence shorter average generation time, this disease will have a quicker initial growth. So, eventhough one has quicker growth that the other, they will eventually result in the same final size (approximately of course).

The above example illustrates that some additional information, beside the initial growth rate, is needed in order to infer $R_0$ and the final fraction getting infected $r(\infty)$. The needed quantity is the so called \emph{generation time distribution} $g(s)$, which quantifies the distribution of the time between getting infected and infecting a new individual (cf.\ \cite{WL07} and \cite{Sven2007}). Or, equivalently, an individual infects new individuals at average rate $R_0g(s)$ $s$ time units after infection. For the standard stochastic SIR epidemic $g(s)=P(I>s)/E(I)$ but the generation time distribution can be computed for more realistic models allowing for latent periods and time varying infectivity. Using theory for branching processes (e.g.\ \cite{Jage75}) it is well-known that, given the generation time distribution $g(s)$, the exponential growth $\rho$ and the basic reproduction number $R_0$ are connected to each other through the Lotka equation
$$
\int_0^\infty e^{-\rho t}g(t)dt=\frac{1}{R_0}.
$$
So, if we observe the emerging phase we can estimate $\rho$, which together with knowledge about the generation time distribution will give us an estimate of $R_0$, and hence of the final size using the theory of Section \ref{Sec-asymptotics}. It remains to get an estimate of the generation time distribution $g(\cdot )$.

To estimate the generation time distribution is however often quite hard, in particular for an emerging outbreak for which there might not be much historical information. Methods for doing this often rely on contact tracing and comparing the onsets of symptom of cases and their likely infector. We refer to Team WER et al.\ (2014 ??) for a recent treatize on such estimates for the Ebola outbreak. \cite{BritScal2018} high-light some specific difficulties with such estimation problems, which could lead to biased estimates of $R_0$: early in an outbreak short generation times will be over represented, if individuals having multiple potential infectors are neglected will make remaining generation systematically shorter, and the random delay between infection and onset of symptoms can make generation times estimated with too high variance. All three effects lead to $R_0$ being \emph{underestimated} if not adjusted for.

\subsection{Inference based on endemic levels}

In Section \ref{Sec_endemic} the endemic levels $(\tilde s, \tilde i,\tilde r)$ of susceptibles, infectives and recovered (=immune), for so-called childhood diseases giving life long immunity, were given in Equation (\ref{end-level}). If we observe a community at endemicity we can therefore estimate $R_0$ simply by
$$
\hat R_0^{(endemic)}=\frac{1}{\tilde s}.
$$
A probabilistic analysis of the endemic model is much harder than the model in a fixed and closed community. For this reason there are currently no available plug-in estimates of the standard error of this estimate. However, we can say a bit more about the estimate itself.

At first it might not seem that easy to obtain the data observation $\tilde s$, the fraction susceptible at endemicity. But, since we are considering diseases giving life-long immunity, the length of the susceptible life-period of an individual is identical to the age at which he/she gets infected. Since we are considering a community at equilibrium the fraction of individuals being susceptible will therefore equal the average relative part of a life an individual is susceptible, and this is simply  the average age of infection $a$ divided by the average life-length $\ell$: $\tilde s=a/\ell$. Both these numbers are easily obtained: the former from the medical authorities and the latter from national statistics data. 

As an illustration, suppose the average life-length equals 75 years, and the typical age of infection of some disease not currently vaccinated for, is 5 years. Then $\tilde s=5/75=1/15$ which hence implies that $\hat R_0=15$.

\subsection{Inference for extended models}

In Section \ref{Sec-heterogeneities} several extensions of the standard stoachastic SIR epidemics were discussed, bringing in realism in terms of various sorts of heterogeneities. These were for example to acknowledge that individuals are of different types, having different susceptibilities and infectivities between different types, for example due to age, gender and/or prior history to the disease; models which are often referred to as multitype epidemic models. Another heterogeneity lies in how people mix with each other; if for example considering influenza, including household structure into the model makes sense, whereas if considering STI's, a network mimicking the network of sex-contacts is more relevant. Finally, there might be heterogeneity in infectivity over time, either calendar time because of seasonal differences and/or time since infection where infectivity may first increase, then peak, followed by a slow decay down to zero.

To make inference in such more complicated situations, including also other aspects, is what most of the forthcoming chapters are dealing with. We  hence refer to later sections for such statistical analyses except giving a few qualitative statements.

If observing the final outcome of a multitype epidemic the fraction infected in each type is observed, and it is assumed that the community fraction of the different types are known. If there are $k$ types of individuals, the data vector is hence $k$-dimensional. However, the number of parameters is greater than $k$, whether assuming a completely general contact matrix between different types (having dimesion $k^2$) or assuming separable mixing where the contact rate between two types is the infectivity of the infective type multiplied by the susceptibility of the receiving type (dimension $2k$). As a consequence, it is not possible to estimate all model parameters consistently, and what is worse, it is not even possible to estimate $R_0$ consistently. Without additional knowledge, all that is possible to do is to give a range of possible values of $R_0$ (cf.\ \cite{Brit1998}).

When it comes to household models, it is possible to estimate the transmission rates both within and between households whether observing temporal or final size data. Intuitively, the more cases are clustered in certain households the more spreading there is within households. From these estimates it is possible estimate $R_0$, or rather another threshold parameter $R_*$ called the household reproduction number, cf.\ \cite{Ball97}.

Network epidemic models, and inference for such, have received much attention in the literature during the last two decades. From an inference point of view, the statistical methodology differs whether the network is observed globally, locally or not at all, beside observing infected individuals. If the complete network is observed, inference is quite straightforward: susceptible individuals are exposed by infectious neighbours, and by observing when infection takes place and how long infectious periods last it is possible to infer disease model parameters. If the network is only observed locally, e.g.\ the number of neighbours of infected individuals, or the more common situation that the underlying network is not observed at all, expect possibly some summary statistics such as mean degree and/or clustering, then inference becomes much harder. Individuals that get infected are usually unrepresentative in having many neighbours thus exposing themselves to higher risk of transmission, and it is not observed which are the underlying links responsible for infection, making estimation of $R_0$ impossible without additional assumptions.

The final type of heterogeneity regards variation in either calendar time or time since infection. Varying infectivity due to calendar time is often referred to as seasonality and is usually modelled by a sinodal curve. It is possible to include such a function and to estimate parameters using e.g.\ reported incidence over the year. As for the infectivity function as a function of time since infection, denoted the generation time distribution,  is often estimated from contact tracing, see e.g.\ \cite{WHO2014}. But as mentioned in Section \ref{Sec-emerging} this is often associated with potential risk for biases.

\section{Introducing prevention: modelling and inference} \label{Sec_prevention}

One of the main reasons for modelling and making inference for epidemics is to better understand them, and in particular to understand what preventive measures are needed to reduce or preferably completely stop an outbreak. In the current section we focus on the preventive measures which make susceptible individual no longer at risk of infection. This can be acheived in different ways depending on the application: an individual may get vaccinated, isolated or for STIs stop being sexually active or only having safe sex. In what follows we use the term \emph{vaccination} but bear in mind that this may have alternative meanings.

Suppose that a fraction $v$ of the community is vaccinated prior to the arrival of the outbreak, or, in the endemic setting, suppose that a fraction $v$ of all new-born individuals are vaccinated. Further,  assume that the vaccine gives 100\% protection (there also exist model extensions allowing for partial \emph{vaccine efficacy}). The basic reproduction number is then reduced to $R_v=R_0(1-v)$, since only the fraction $1-v$ of the infectious contacts are with non-vaccinated individuals. As a consequence, there will be no outbreak (or the disease will vanish in the endemic setting) if $R_v\le 1$. But this is equivalent to $v\ge 1-1/R_0$. The value giving exact equality is known as the critical vaccination coverage and denoted $v_C=1-1/R_0$, a very important quantity when aiming at preventing an outbreak or making an endemic disease disappear. 

Because we have estimates of $R_0$ from final size data, an estimate of $v_C$ for the same data is immediate:
\begin{equation}
\hat v_C = 1-\frac{1}{\hat R_0}= 1- \frac{s\tilde r_s}{-\ln(1-\tilde r_s) }.
\end{equation}
Recall that $s$ denotes the initial fraction susceptible in the community in which the outbreak took place, and $\tilde r_s$ the observed fraction infected among the initially susceptibles. A standard error for $\hat v_C$ can be obtained using similar methods as for $\hat R_0$. The result says that 
\begin{equation}
s.e.(\hat v_C )= \frac{1}{\sqrt{ns}}\sqrt{\frac{1+c_v^2(1-\tilde r_s )\hat R_0^2s^2}{ \hat R_0^4 s^2  \tilde r_s (1-\tilde r_s )}},
\end{equation}
where as before, $c_v$ denotes the coefficient of variation of the infectious period, which can be conservatively estimated to 1 if unknown.

For endemic diseases having a fraction $\tilde s$ susceptible, the corresponding estimate of $v_C$ equals 
$$
\hat v_C^{(endemic)}=1- \tilde s.
$$
To obtain a standerd error for this estimate remains an open problem, but the standard error should be of order $1/\sqrt{n}$.

\section{Discussion}

Reality is often complicated, and more realistic models having more complicated inference procedures are many times to be preferred as compared to the simple models of the current chapter. However, a recommendation is to complement such analyses with the simpler methods of the current chapter. If the estimates from the simpler methods are close to the ones in the more complicated models this is reassuring, and if not it is worth spending some time to understand why this is not the case. 

We again stress the importance of acknowledging that not all infected individuals are usually reported, often due to no  or minor symptoms (asymptomatic infections). 

In the current chapter we did not consider estimation of vaccine efficacy, usually inferred in a clinical trial in which certain individuals are vaccinated and others not. In fact there are several different vaccine efficacies: in terms of susceptibility, symptoms, infectivity if infected, and others. This rather complicated inference problem is investigsted in detail in \cite{HLS2010}

One heterogeneous feature which was not considered in the current chapter were spatial aspects, where most likely, the risk of transmitting someone decrease with the distance between the steady locations of the two individuals (particularly relevant in wild-life and plant populations). 

We end by giving a general rule of thumb: various heterogeneities play a bigger role the less transmittable the disease is, So homogeneous mixing models often work satisfactorily for measles and similar childhood diseases, but various heterogeneities need to be included when analysing e.g.\ STI outbreaks.


\end{document}